\documentclass{aa}

\usepackage{times}
\usepackage{amssymb}
\usepackage{epstopdf}
\usepackage{url}
\usepackage{natbib}
\bibpunct{(}{)}{;}{a}{}{,}
\usepackage{graphicx}
\usepackage{amsmath}

\def\ergscm2 {erg\,s$^{-1}$cm$^{-2}$}

\def\cm2 {cm$^{-2}$}

\begin{document}
        
\title{The variable absorption in the X-ray spectrum of GRB~190114C}

\titlerunning{Variable absorption of GRB~190114C}

\author{Sergio Campana\inst{1}, Davide Lazzati\inst{2}, Rosalba Perna\inst{3,4}, Maria Grazia Bernardini\inst{1}, Lara Nava\inst{1}}

\authorrunning{S. Campana et al.}

\institute{
{INAF-Osservatorio Astronomico di Brera, Via E. Bianchi 46, I-23807 Merate (LC), Italy}
\and
{Department of Physics, Oregon State University, 301 Weniger Hall, Corvallis, OR 97331, USA}
\and
{Department of Physics and Astronomy, Stony Brook University, Stony Brook, NY 11794-3800, USA}
\and
{Center for Computational Astrophysics, Flatiron Institute, New York, NY 10010, USA}
}

\abstract{
GRB~190114C was a bright burst that occurred in the local Universe ($z=0.425$). It was the first gamma-ray burst (GRB) ever detected at TeV energies, thanks to MAGIC. We characterize the ambient medium properties of the host galaxy through the study of the absorbing X-ray column density. Joining {\it Swift}, {\it XMM-Newton}, and {\it NuSTAR} observations, we find that the GRB X-ray spectrum is characterized by a high column density that is well in excess of the expected Milky Way value and decreases, by a factor of $\sim 2$, around $\sim 10^5$~s. Such a variability is not common in GRBs.
The most straightforward interpretation of the variability in terms of photoionization of the ambient medium is not able to account for the  decrease at such late times, when the source flux is less intense. Instead, we interpret the decrease as due to a clumped absorber, denser along the line of sight and surrounded by lower-density gas. 
After the detection at TeV energies of GRB~190114C, two other GRBs were promptly detected. They share a high value of the intrinsic column density and there are hints for a decrease of the column density, too. We speculate that a high local column density might be a common ingredient for TeV-detected GRBs.}

\keywords{Gamma-ray burst: general -- dust, extinction -- Gamma-ray burst: individual: GRB 190114C}

\maketitle

\section{Introduction}

Since the early observations of
Gamma-Ray Bursts (GRBs), it has been recognized that the study of their immediate environment plays a very important role in order to gain a more complete understanding of the properties of their progenitors (e.g. \citealt{Mirabal2003,Fryer2006,Delia2007,Delia2009,Hjorth2012,Kruler2015,Vergani2015,Perley2016}). 

For example, low density environments, as typical of the outer regions of galaxies, are expected preferentially in association with long-lived progenitors, such as compact object binaries, due to their generally long merging times (e.g. \citealt{Berger2005,Belczynski2006}). This is characteristic of short GRBs.
On the other hand, GRBs associated with the collapse of massive stars (such as the long GRBs, \citealt{Stanek2003,Hjorth2003}) are expected to occur within star forming regions, thus reflected in higher densities of the circumburst medium. Even more informative, when it can be inferred from the data, is the density profile in the immediate vicinity of the source, since this carries information on the last stages of the life of a massive star, such as the extent and amount of mass lost in a wind, or the presence of shells of material ejected prior to the GRB \citep{Racusin2008}, as expected in a two-step explosion \citep{Vietri1998}.   

Several proposals for the origin of the X-ray absorption in excess of the Galactic value in GRBs have been suggested, both in
individual sources as well as
from a sample perspective. The easiest explanation involves material in the host galaxy, either local to the GRB \citep{Galama2001,Stratta2004,Gendre2006,Schady2007,Campana2012} or on a more extended basis (HII regions, Watson et al. \citeyear{Watson2013}; molecular clouds, Reichart \& Price \citeyear{Reichart2002}; Campana et al. \citeyear{Campana2006}).
Another possibility involves the metal contribution from the intergalactic medium (IGM) along the line of site \citep{Behar2011,STarling2013,Campana2015,Dalton2020}. These possibilities are not mutually exclusive.

One way to probe the properties of the absorbing medium is via time-resolved measurements of the column density of the absorbing material along the line of sight from the observer to the source, $N_{\rm H}$. If time-variability in the absorption is observed, it can yield powerful information on the immediate environment of the source.

The long gamma-ray burst GRB~190114C, triggered on January 14, 2019 by both the Neil Gehrels {\it Swift} Observatory and the {\it Fermi} satellite \citep{Gropp2019,Ajello2020}, was followed by an extensive observational campaign at various wavelengths, due to its high luminosity ($E_{\gamma,\rm iso}=(2.5\pm 0.1)\times 10^{53}$~erg) and relatively low redshift ($z=0.425$, \citealt{Selsing2019}), making the burst extremely bright \citep{GRB190114Ca}. The closeness of the GRB makes the IGM contribution negligible (e.g. Arcodia et al. \citeyear{Arcodia2018}). In addition, the optical absorption is high $E(B-V)\sim0.83$, supporting the evidence for a high intrinsic absorption \citep{deUgartePostigo2020}.
Remarkably, observations by the MAGIC collaboration also revealed teraelectronvolt (TeV) emission, the first time from a GRB \citep{GRB1901114}. 

Here we present X-ray observations of GRB~190114C with {\em Swift}/XRT, from the earliest observation time window (starting 68~s after the burst) to the latest afterglow observations on a timescale of $\sim 10$~days, together with {\em XMM-Newton} and {\em NuSTAR} long observations. During this period, the effective column density $N_{\rm H}$ was observed to decline by approximately a factor of two. 

The most natural explanation for a decline in $N_{\rm H}$ during the time that the source is active is photoionization of the medium along the line of sight to the observer by the strong X-ray/UV radiation accompanying the burst. 
As the radiation propagates into the medium, it gradually photoionizes it, hence reducing the opacity of the medium encountered by the later radiation front. This phenomenon can result in  time-dependent absorption lines (and has been
discussed in both the optical, \citealt{Perna1998,Mirabal2002,Dessauges2006,Thoene2011}, and the X-rays, \citealt{Bottcher1999,Lazzati2001}), as well as more generally for the effective column density $N_{\rm H}$, which is an integral quantity typically measured in X-ray spectral fits \citep{Frontera2004,Campana2007,Grupe2010}. 
Whether the variability is appreciable enough to be measurable in time-resolved spectra depends on both the brightness of the source, as well as on the radial extent of the absorbing medium. GRB~190114C fulfils these conditions and, with its measured variability in $N_{\rm H}$, offers us an opportunity to probe the environment in the immediate vicinity of the source.

This paper is organized as follows: Sec.~2 presents the X-ray data and the time-resolved spectral fits, leading to the measurements of the variable $N_{\rm H}$.
In Sec.~3 we perform a detailed statistical analysis aimed at determining whether time-dependent photoionization of the line-of-sight absorbing medium can provide a reasonable fit to the data for $N_{\rm H}(t)$. The large $\chi^2$ of the best fit leads us to conclude that this is not a likely explanation for the observed variability. In Sec.~4 we suggest some alternative explanations. In particular, we show that an absorber with a low surface filling-fraction can reproduce the observed variability, at least qualitatively. We summarize and conclude our work in Sec.~5.

\section{X-ray data and light curve modeling}

The Neil Gehrels {\it Swift} Observatory (Gehrels et al. 2004), after the detection of the GRB with the Burst Alert Telescope (BAT; \citealt{Gropp2019}), autonomously re-pointed its narrow field instruments and the X--ray Telescope (XRT) started observing 68 s after the event. When XRT started observing, BAT was still collecting useful data, allowing for a 130 s superposition.
{\it Swift}/XRT observed the GRB for about a month with 26 exposures. Useful spectral data stop at $\sim 11$ d. 
Together with {\it Swift}, GRB 190114C has been observed once by {\it NuSTAR} and twice by {\it XMM-Newton}. A log of all the X-ray observations can be found in Table \ref{logxray}.

\begin{table}[]
\centering
\begin{tabular}{cccc}
\hline
Telescope        & Obs. ID.    & Start time     & Expos.\\
                 &             & (2019-)        & (s)\\
\hline
{\it Swift}/XRT  & 00883832000 & 01-14 20:39:00 & 	558.9\\
{\it Swift}/XRT  & 00883832001 & 01-14 21:55:36 & 	19412.5\\
{\it Swift}/XRT  & 00883832002 & 01-15 14:18:35 & 	7454.2\\
{\it Swift}/XRT  & 00883832003 & 01-16 02:44:35 & 	2788.1\\
{\it Swift}/XRT  & 00883832004 & 01-16 09:10:34 & 	2419.5\\
{\it Swift}/XRT  & 00883832005 & 01-17 05:57:36 & 	2392.0\\
{\it Swift}/XRT  & 00883832006 & 01-17 09:06:35 & 	2725.4\\
{\it Swift}/XRT  & 00883832007 & 01-18 05:45:34 & 	4974.5\\
{\it Swift}/XRT  & 00883832008 & 01-19 12:05:35 & 	3582.9\\
{\it Swift}/XRT  & 00883832009 & 01-20 13:31:34 & 	4681.1\\
{\it Swift}/XRT  & 00883832010 & 01-21 16:46:35 & 	3066.4\\
{\it Swift}/XRT  & 00883832012 & 01-22 02:19:51 & 	4979.5\\
\hline
{\it XMM-Newton} & 0729161101  & 01-15 05:12:01 & 35421.9\\
{\it XMM-Newton} & 0729161201  & 01-24 06:17:08 & 10838.0\\
\hline
{\it NuSTAR}     & 90501602002 & 01-15 19:16:09 & 52391.2\\
\hline
\end{tabular}
\caption{Log of the X--ray observations.}
\label{logxray}
\end{table}

\begin{table*}[]
\centering
\begin{tabular}{ccc|cc}
\hline
Number & Time Slice & Instrument & $N_{\rm H}(z)$ & $E_{\rm peak}$\\
 & (s)        &            & ($10^{22}$ cm$^{-2}$) & (keV) \\
\hline
1 & 67.7--132     & BAT - XRT (WT grade 0)& $8.79^{+0.57}_{-0.68}$& $15.2^{+13.7}_{-10.5}$\\
2 & 132--196      & BAT - XRT (WT grade 0)&$8.94^{+0.75}_{-0.59}$ &$41.5^{+53.2}_{-22.1}$\\
3 & 196--644      & XRT (WT grade 0)      &$8.06^{+0.39}_{-0.65}$& $>37.5$\\
4 & 3844-4164     & XRT (WT grade 0)      & $8.86^{+0.80}_{-1.12}$&$<6.7$\\
5 & 5507--9988   & XRT (PC)              &$8.02^{+1.51}_{-1.23}$&$>4.7$\\ 
6 & 9988--29315  & XRT (PC)              &$7.38^{+0.77}_{-0.76}$&$>8.2$ \\
7 & 32515--80900  & XMM (pn) - XRT (PC)   &$6.50^{+0.17}_{-0.26}$& $13.8^{+14.6}_{-4.3}$ \\
8 & 80644--142660 & NuSTAR (A+B) - XRT (PC)&$7.75^{+1.33}_{-1.19}$& $7.1^{+7.2}_{-5.6}$ \\
9 & 205508--769284& XRT (PC)              &$4.25^{+2.41}_{-1.27}$&$-$ \\
10 & 811350--855900& XMM (pn+MOS1+MOS2)    &$4.26^{+0.72}_{-0.83}$&$ 4.3^{+0.6}_{-0.5}$ \\
\hline
\end{tabular}
\caption{Time slices used for spectral analysis. Errors are at $90\%$ confidence level for one parameter of interest.}
\label{timeslice}
\end{table*}

{\it Swift}/XRT collected data in Windowed Timing (WT) mode up to $\sim 5,000$ s from the burst, testifying for the burst brightness. The spectra from these WT data were extracted selecting single pixel events only, in order to increase the spectral resolution and minimise the effects of charge redistribution at low energies. We also used the new analysis software included in {\tt HEASOFT} v.6.26 to minimise the effects of charge trapping. These data were extracted manually. 
To extend the XRT spectral range, we also extracted BAT data up to $\sim 200$ s. We analysed {\it Swift}/BAT data and processed them with the standard {\it Swift} analysis software included in {\tt HEASOFT} v.6.26 and the relevant calibration files. We extracted $15-150$ keV BAT spectra and response matrices with the \texttt{batbinevt} and \texttt{batdrmgen} tasks in FTOOLS in two time intervals: $68-132$ s and $132-196$ s. Beyond this time there was no sufficient signal to extract a spectrum.

All the other XRT data were collected in Photon Counting (PC)  mode. These XRT spectra were extracted using the Leicester University tools\footnote{https://www.swift.ac.uk/xrt\_spectra/}, correcting automatically for all the instrumental features \citep{Evans2009}.
For all the spectra, we retained data in the 0.3--10 keV energy range and standard grade selection (1-12). We binned the spectra to 1 photon per energy channel to allow for the use of C-statistics.

The first {\it XMM-Newton} observation started 8.3 hr after the burst. Data were processed following standard filtering criteria according to {\it XMM-Newton} threads\footnote{https://www.cosmos.esa.int/web/xmm-newton/sas-threads}. The net exposure time is 35 ks. Given the high count rate, we retained the pn data only. Source data were extracted from a 800-pixel region, the background from a close-by region free of sources and of 1000 pixel radius. The second observation started 9.4 d after the event. The count rate is lower by a factor of $\sim 100$, so we retained all the three EPIC cameras. Due to the high background, we filtered the data with a pn threshold of 0.8 counts s$^{-1}$. The net exposures are 10.8, 9.3, and 9.3 ks for the pn, MOS1, and MOS2, respectively. Being the source weaker, we extracted photons from a 400 pixel radius region. We retained data in the 0.3--10 keV energy range. We binned the data to 1 photon per energy channel.

\begin{figure}[htb]
    \centering
     \includegraphics[width=1.05\columnwidth]{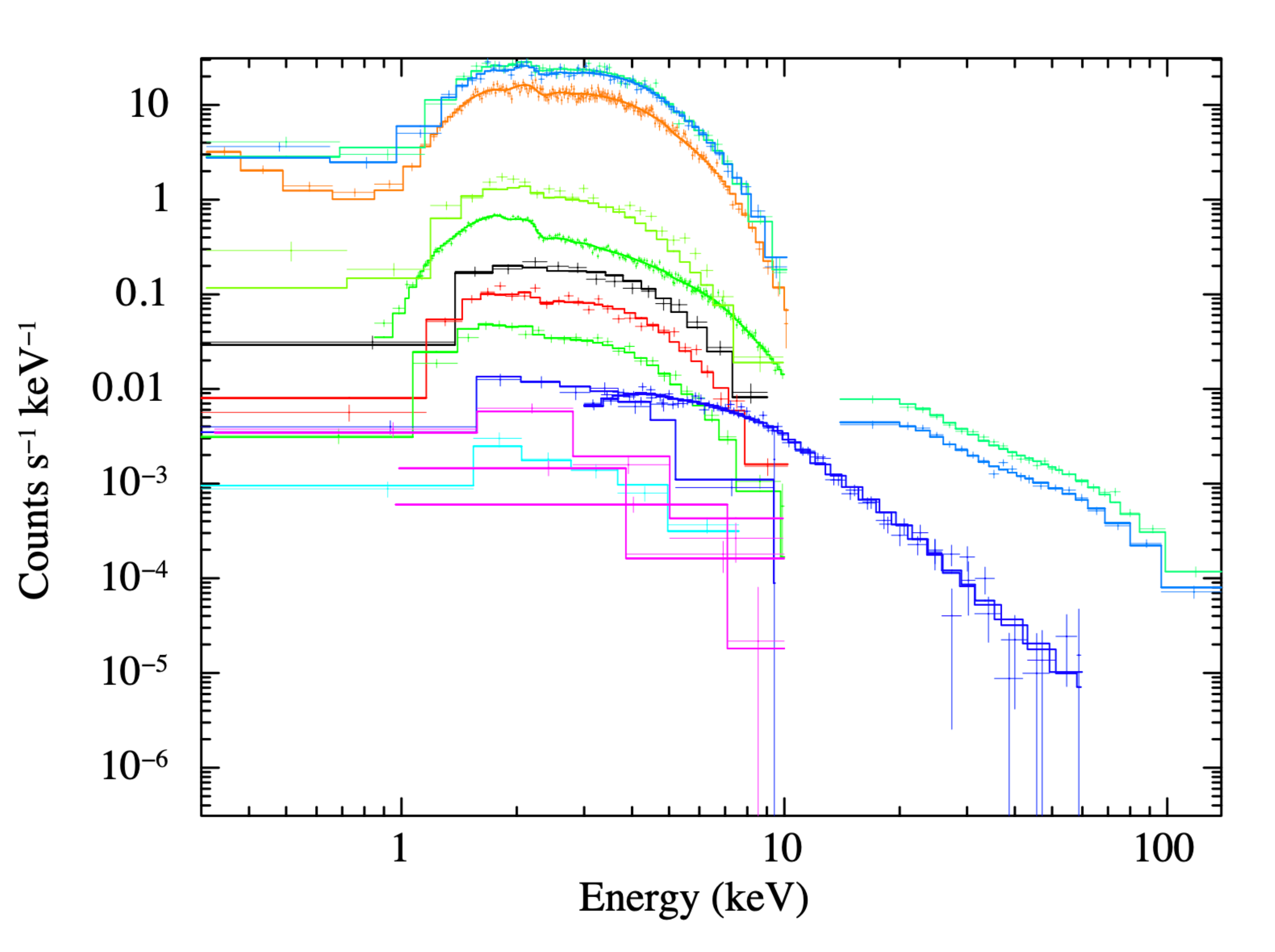}
    \caption{Fit of the spectral data. Data were heavily rebinned for displaying purposes. From top to bottom, spectra are in order of increasing time and decreasing flux. Spectra as numbered in Table \ref{timeslice}, are plotted in green+cyan (1), blue+cyan (2), orange (3), yellow+green (4), black (5), red (6), green (7), blue (8), light blue (9), and magenta (10).}
    \label{fig:spettri}
\end{figure}

The {\it NuSTAR} observation took place 22.4 hr after the burst. We started from the standard cleaned data, grouping the three event files for each instrument. We extracted the spectra from a region of $49''$ radius centered on source. The background was extracted from a circular region of $109''$ radius. We considered data in the 3--60 keV energy range and binned the data to 1 photon per energy channel.

We divided the data into 10 time slices (see Table \ref{timeslice}). 
Spectra were fitted with {\tt XSPEC} (v12.11.1) and we adopted C-statistics \citep{Cash1979}. Although the collected data span a large time interval (68 s - 10 d) and we are dealing with a large wealth of data, we adopted a simplistic spectral model: two smoothly joined power laws ({\tt sbpl} in {\tt XSPEC}) with Galactic and intrinsic absorption, modelled with {\tt tbabs}. The Galactic absorption was fixed to $7.45\times 10^{19}$ cm$^{-2}$ \citep{Willingale2013}. The intrinsic column density was evaluated at the GRB redshift ($z=0.425$) and left free to vary from one time slice to another. The two power law indices were tied together among all observations and only the peak energy was free to vary. The smoothness parameter was kept fixed to 1 and we verified this is a parameter that cannot be fitted and results do not vary too much for other different choices. A constant factor was added to cope with instrument calibration uncertainties resulting in different values of the power law normalisation. The same value of the constant was adopted for the same instrument. A value of 1 was kept for {\it Swift}/XRT data taken in PC mode.

The overall fit is good with a C-statistic of 6293.2 with 6850 degrees of freedom.  The normalization constants are close to one, as expected (see Fig.~\ref{fig:spettri}).
The power law photon indices are $\alpha_1=1.47_{-0.30}^{+0.18}$ and $\alpha_2=2.14_{-0.08}^{+0.19}$ (errors were computed for $\Delta\chi^2=2.71$).
These values of the photon indices are consistent with synchrotron radiation from a non-thermal electron population injected with spectral index $p\sim2.2$. The intrinsic column density and the peak energy decrease with time (see Table \ref{timeslice}). The decreasing peak energy (corresponding to the cooling energy) is suggestive of a constant density medium \citep{Panaitescu2000}. The initial intrinsic column density is very high ($\sim 9\times 10^{23}$ cm$^{-2}$, where the mean intrinsic column density for bright bursts is $\sim 5\times 10^{21}$ cm$^{-2}$, \citealt{Campana2012}).

\begin{figure}[htb]
    \centering
    \includegraphics[width=\columnwidth]{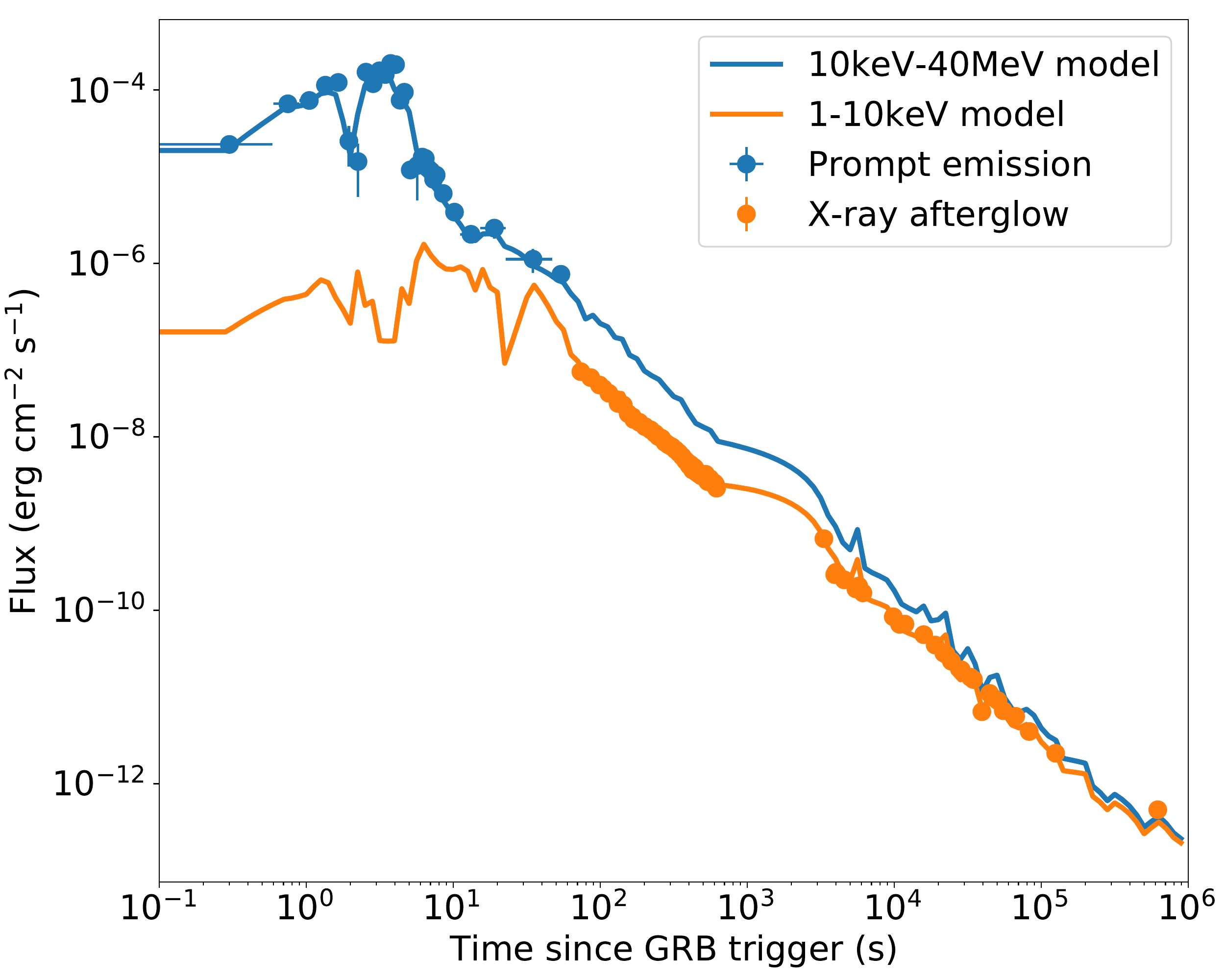}
    \caption{Comparison between the prompt and X-ray afterglow observations of GRB~190114C with the model light curve and spectral evolution adopted in the 
    photoionization code. Prompt emission data from Ravasio et al. (\citeyear{Ravasio2019}); afterglow data from this work.}
    \label{fig:alum}
\end{figure}

We extracted the GRB light curve from the {\it Swift}/XRT light curve repository and converted into 1--10 keV flux (Fig.~\ref{fig:alum}). For each spectral slice we computed the (variable) conversion factor from XRT count rate to 1--10 keV unabsorbed flux. Then, we interpolated these conversion factors and converted all the count rates to the flux light curve shown in Fig.~\ref{fig:alum}.

To reconstruct the 1--10 keV light curve before 70 seconds, we used the flux measured in the $10$ keV $-$ $40$ MeV band by Ravasio et al. (\citeyear{Ravasio2019}). The spectrum is assumed to be either a broken power-law or a single power-law, as requested by the fit. 

As apparent from Table~2, the column density $N_{\rm H}$ decreases with time.
A linear fit yields a $\chi^2=81.6$ with 8 degrees of freedom, hence ruling out a linear decrease and calling for a different functional form for the measured variability. A possible fit involves a power law (with an index of 0.22) and a constant, yielding a $\chi^2=10.0$ with 7 degrees of freedom. In the following, we explore physical models which can lead to a time-decreasing $N_{\rm H}$.

\section{Modeling with time-dependent photo-absorption}

The most natural explanation for the observed reduction in the column density to the source during the time of observation is photoionization by the strong UV/X-rays radiation \citep{Perna2002,Lazzati2001,Lazzati2002}. We thus here quantitatively explore whether this interpretation is supported by the data. 

The GRB radiation source is assumed to turn on in a medium in thermal equilibrium at an initial temperature of $T_{\rm in}=10^3$~K. 
The time-dependent photoionization of the medium is computed using the code developed by \citet{Perna2002,Perna2003}, and described in detail in those papers. 
The code includes 13 elements, that is Hydrogen and the 12 most abundant astrophysical elements: He, C, N, O, Ne, Mg, Si, S, Ar, Ca, Fe, Ni, with solar  abundances. The ionic number densities of those elements are computed as a function of space and time as the flux from the source propagates. 

In order to apply the output of the code to the X-ray data, we need to transform the output to be readily compared to $N_{\rm H}$ under the assumption that all the material is cold (i.e. neutral; see \citealt{Morrison1983}). 
Within our formalism, this is equivalent to the approximation that the time and frequency-dependent optical depth can be decomposed as $\tau(\nu,t)=N_{\rm H}\sigma(\nu)$, where $\sigma(\nu)$ is the average cross section at frequency $\nu$ weighed by the element abundance and it is assumed to be independent of time, i.e. independent of the ionization state of the elements\footnote{Note that this approximation is equivalent to assuming that the composition and temperature of the absorbing gas does not change with time. The same assumption was made in fitting the data with the {\tt ztbabs} package for X-ray absorption.}. 
We can then write the time-dependent column density 
in each observation band $[E_1,\, E_2]$ as
\begin{equation}
N_{\rm H} = N_{\rm H} (0) \left<
\frac{\tau(\nu,t)}{\tau (\nu,0)}
\right>_{[E_1,E_2]}\,, 
\label{eq:NH}    
\end{equation}
where the brackets indicate the average over the corresponding energy band.

We consider two types of environments, a wind (which may be expected for massive stars) and a uniform medium, the latter 
distributed between a minimum and a maximum radius. We vary the latter in the range $10^{18} {\rm cm}\leq R_{\rm max}\leq 10^{21} {\rm cm}$. For each value of $R_{\rm max}$, we then vary the minimum radius in the range $0.05\leq R_{\rm max}/R_{\rm min}\leq 0.95$. 

\begin{figure}[htb]
\includegraphics[width=1.0\linewidth]{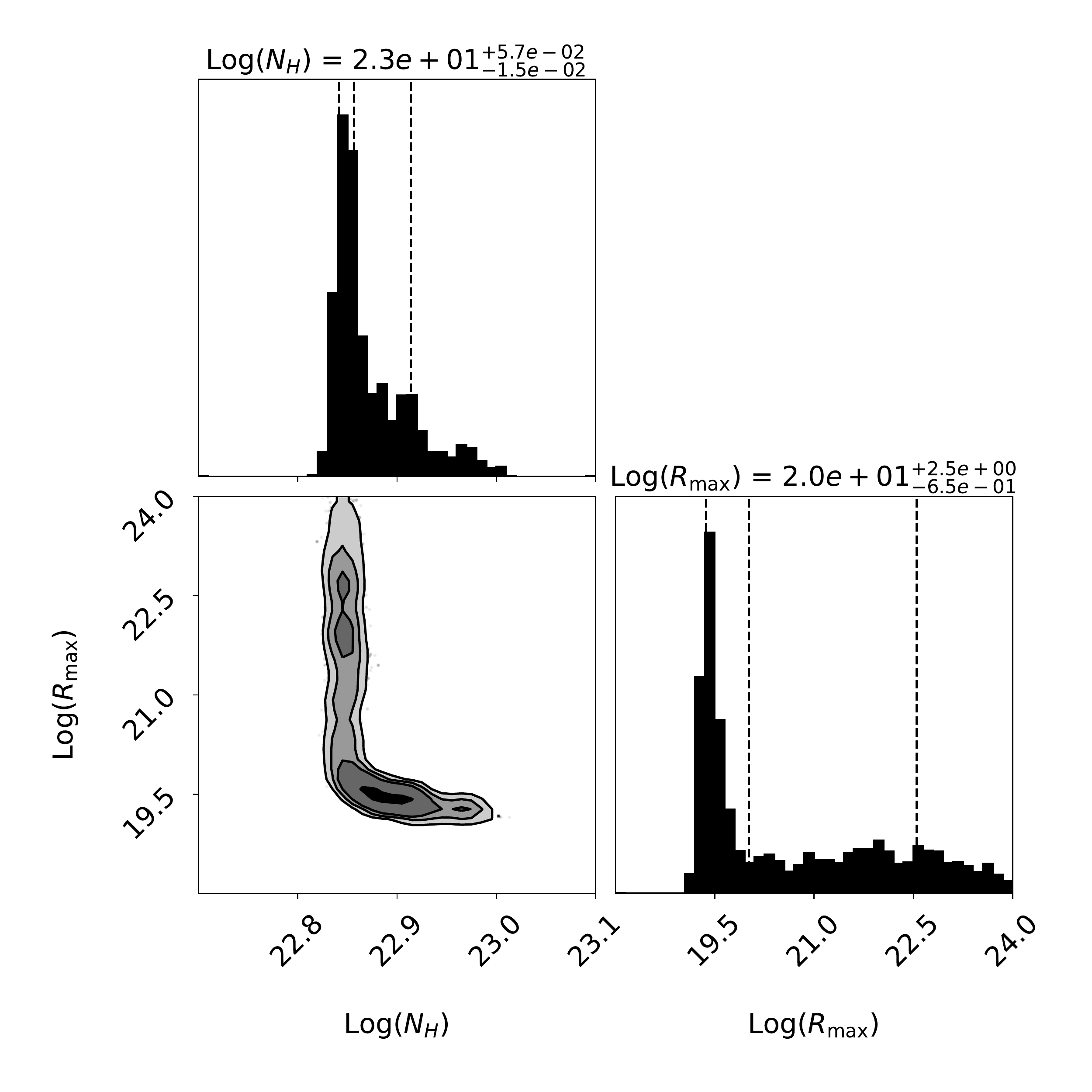}\\
\includegraphics[width=1.0\linewidth]{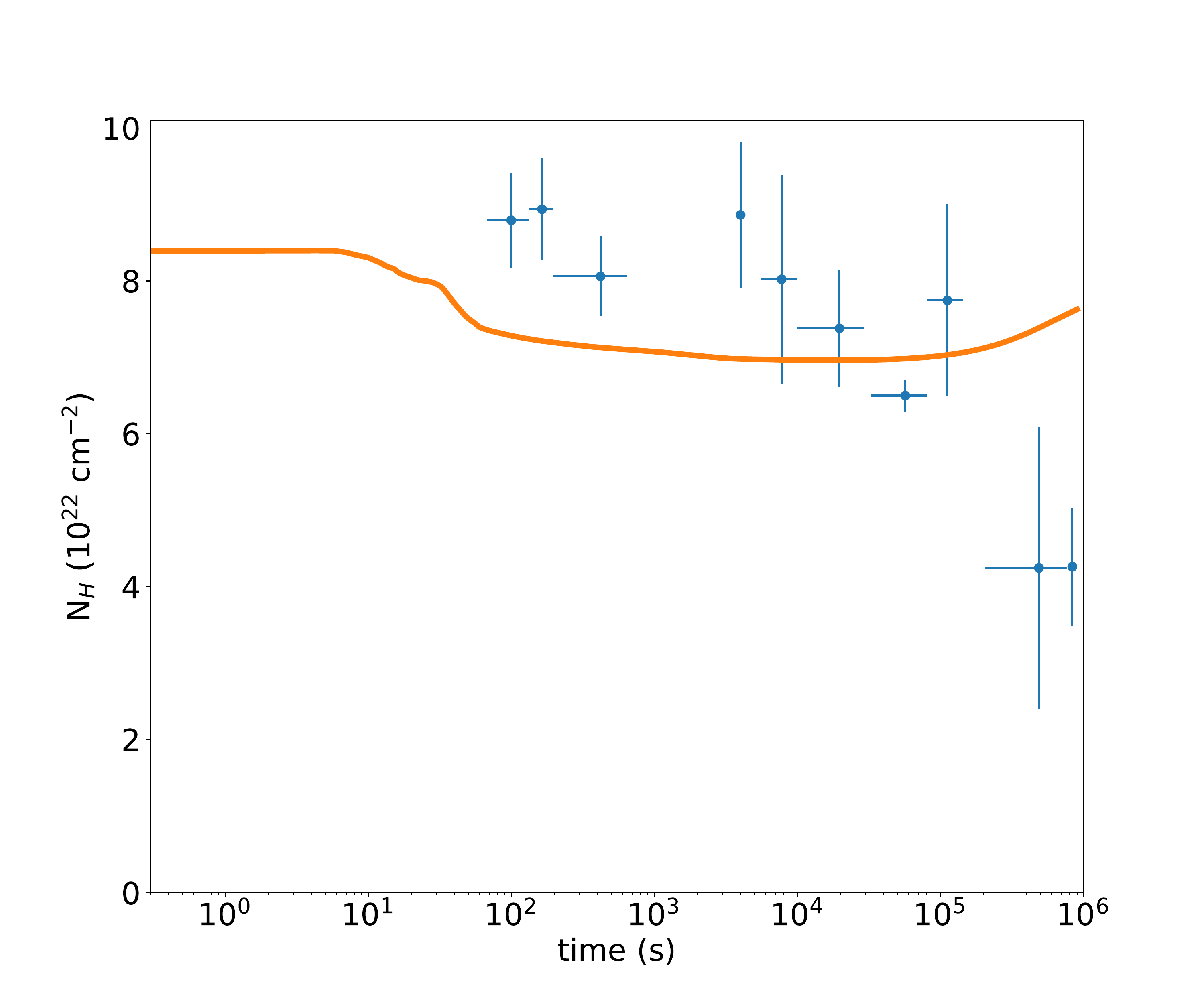}
\caption{Top: Results from the MCMC simulation for modeling via time-dependent photo-absorption. Bottom: Corresponding best fit model from photo-absorption. The large $\chi^2_{\rm red}\sim 5$ shows that time-dependent absorption is not a good explanation to the observed time-variable absorption.}
\label{fig:photoabs}
\end{figure}

A wind-like density profile is described by $n(r) = n_0 (r/r_0)^{-2}$.
From a few test runs, we were able to immediately establish that a wind profile, even with extreme values of the initial column density $N_{\rm H}(0)=10^{25}$~cm$^{-2}$ is ruled out by the data, since in a wind most of the absorbing material is concentrated close to the source, and hence a sizable change in $N_{\rm H}$ happens on a very short timescale. Therefore we focused our analysis on a uniform medium. The set up is such that it includes both the cases of a typical, extended interstellar medium, as well as that of a thin shell, as envisaged in some models for GRB progenitors \citep{Vietri1998}. 
For each combination of radii, the number density of neutral material prior to the burst onset is
then given by $n= N_{\rm H}(0)/(R_{\rm max}- R_{\rm min})$. 

We performed an MCMC analysis to determine the preferred values of $N_{\rm H}(t=0)$ and $R_{\rm max}$. The results are shown in the upper panel of Fig.~\ref{fig:photoabs}. The corresponding best fit is displayed in the bottom panel.
As visually evident, and formalized by a reduced $\chi_{\rm red}^2\sim 5$ (with 8 degrees of freedom), a photoionization model is not a good explanation for the observed column density variability. The L-shaped isocontours in the corner plot reflect the fact that a constant value of absorption (vertical part of the "L": $N_{\rm{H}}\sim7\times10^{22}$~cm$^{-2}$, $R_{\max}>5\times10^{19}$~cm) provide only a slightly worst fit than the case of a higher initial column at lower radii that is progressively eroded by photoionization (horizontal part of the L, shown also in the bottom panel of Fig.~\ref{fig:photoabs}).
From a physical point of view, this result can be understood in light of the fact that photoionization is most effective during the early times, when the source is brighter. Hence the model predicts an early decline, while the data show it at later times, when the flux is much weaker. Additional strain with the data is caused by the fact that a late decay of the column density is better explained if the absorber is confined within a very thin shell. At the same time, such configuration implies a very high density medium, in which recombination is relatively fast. Recombination of free electrons onto ions is, as a matter of fact, the reason for the increase of column density observed at late times in Fig.~\ref{fig:photoabs}. We conclude that time-dependent photo-absorption is not a good explanation for the observed $N_{\rm H}$ variability. In the following, we discuss some alternative explanations.

The MCMC fit for the time-dependent photoionization model includes any local absorber. We attempted a constant density absorber out to a radius $R_{\rm max}$ as well as a geometrically thin absorber located at a distance $R_{\rm max}$ from the burst. The former would reproduce, for example, the local molecular cloud within which the progenitor was born; the latter would instead include the edge of a cavity formed around the progenitor (e.g., an HII region such as in \citet{Watson2013} or a wind termination shock) as well as a physically unconnected cloud that happens to lie along our line of sight to the burst. We find that the geometrically thin absorber (shown in Fig.~\ref{fig:photoabs}) gives a superior fit with respect to the uniform absorber. However, neither gives an acceptable fit, hence excluding the possibility that time-dependent photoionization of a relatively nearby absorber can explain the measured variable $N_{\rm H}$. While a distant absorber could explain the overall large measured column density, it would not explain the observed decrease of the column density at late times.

\section{Alternative explanations for the observed time-dependent absorption}

Alternative explanations for the drop of the absorbing column at late times ($t\gtrsim10^5$~s) require a real change of the column along the line of sight (irrespective of the gas ionization status). This can be due to a time evolution of the ambient medium, to a change of the line of sight, or a combination of the two. Interestingly, we note that the time of the drop in the absorbing column coincides with the estimated jet break time \citep{Fraija2019}.

\begin{figure}
    \centering
    \includegraphics[width=\columnwidth]{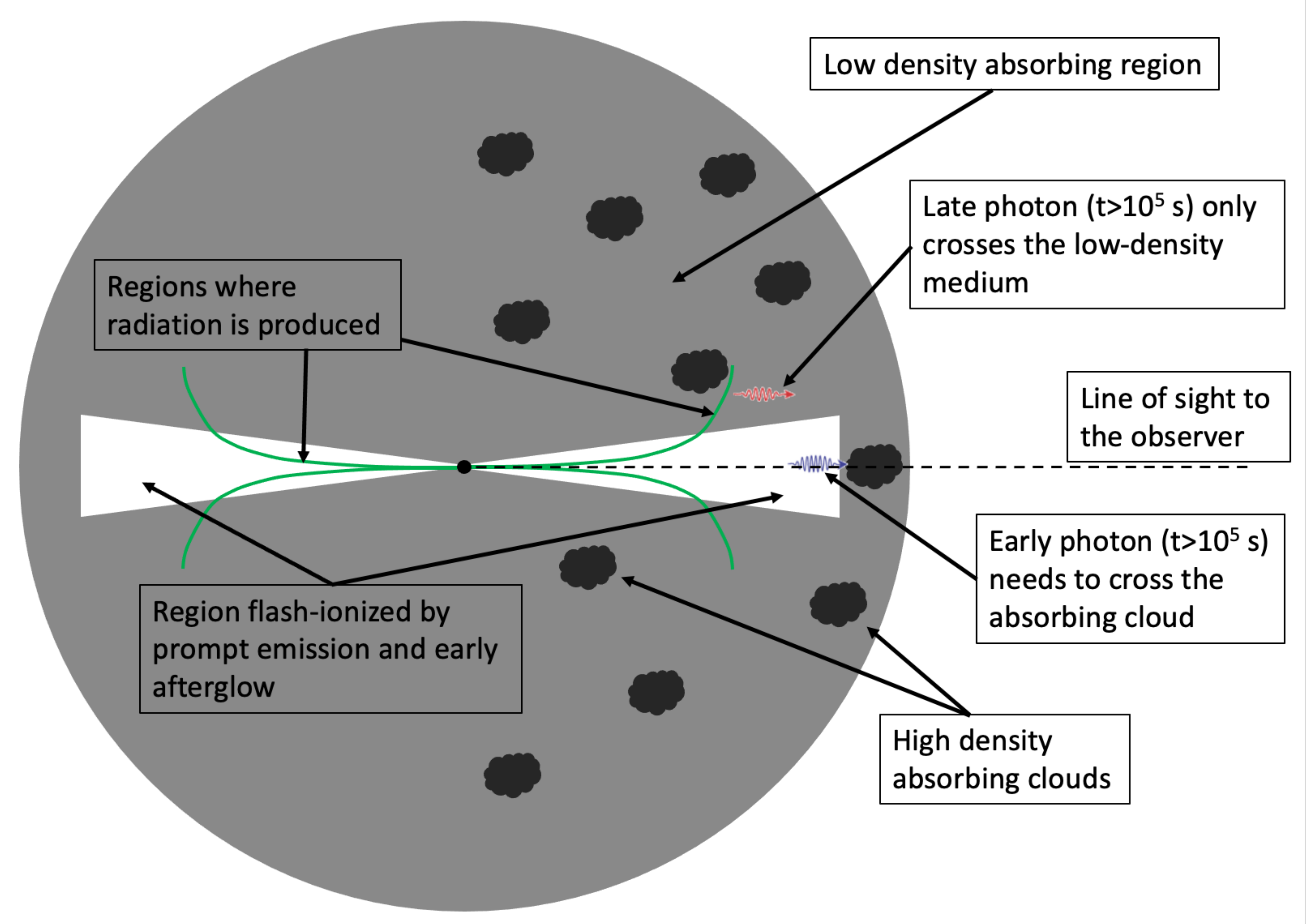}
    \caption{Cartoon of the geometry for explaining a change of absorbing medium coincident with the observed jet break time. Early photons are produced close to the line of sight and need to cross the dense cloud that happens to lie along the line of sight and outside of the flash ionization cone. Late photons are produced at wider angles and travel towards the observer without crossing the high density cloud.}
    \label{fig:cartoon}
\end{figure}

In the first case of a time evolving absorber, we can envisage a medium with clouds or filaments with different densities. It should be kept in mind that the change in absorbing column is of the order of a factor of two and as a consequence a moderate density contrast would suffice. If such clouds move, they might exit the line of sight to the burst at $t\sim 10^5$~s, when the absorbing column drops. 
Given the properties of GRB~190114C, it is possible to estimate the size of the emitting region at $t\sim 10^5$~s as \citep{Derishev2019}:
\begin{equation}
    r_{\perp, \rm{obs}} = R_{\rm{fb}}/\Gamma = \left(
    \frac{6 E_{\rm{iso}}t}{\pi n m_p c \Gamma^4}\right)^{1/4} \sim2.6\times10^{17} \quad {\rm{cm}}
\end{equation}
for a uniform interstellar medium of density $n= 1$~cm$^{-3}$, and using $E_{\rm{iso}}=10^{54}$~erg. The Lorentz factor $\Gamma$ is given by:
\begin{equation}
    \Gamma=\frac12 \left(
    \frac{3 E_{\rm{iso}}}{8\pi n m_p c^5 t^3}\right)^{1/8}\,.
\end{equation}

The absorbing cloud would need to have a size comparable to the emitting region and to have moved a distance comparable to its own size in a time $t_{\rm{obs}}\sim t_{\rm{obs}}\Gamma^2\sim1$~year. Such movement would require a speed comparable to the speed of light ($\sim0.3$~c),  making this model highly unlikely.

Alternatively, one can consider that as time passes the fireball expands and so too does the bright ring from which most of the afterglow radiation is produced \citep{Panaitescu1998}. If the absorber were to be porous, with a $<1$ surface filling factor and a coherence length comparable to the ring size, moderate variations of the absorbing column would be expected. The scenario is sketched in Fig.~\ref{fig:cartoon}. The succession of events would be the following. Initially the fireball flash ionizes a cone within the absorbing cloud of half opening angle equal to the jet angle $\theta_j\sim7.5^\circ$, which we computed from the fireball properties listed above, the estimated jet break time $t_j=10^{5}$~s \citep{Fraija2019}, and assuming a unit density uniform ambient medium. From the flash-ionization runs described above, we know that the cone extends out to approximately $3\times10^{19}$~cm. At the jet break time we observe a steepening in the light curve decay that we know is due to the fact that the fireball has now sideways expanded outside the original opening angle \citep{Rhoads1999}, and therefore outside of the flash-ionized cone. It is therefore possible that from that time onward the radiation crosses an absorber with different properties. For a uniform absorber, we should observe a higher column density for $t>t_j$. In the case of GRB~190114C we observe a drop in the column density. As a consequence, we need to invoke a clumped absorber, with a denser clump along the line of sight surrounded by a lower-density gas (see, sketch in Fig.~\ref{fig:cartoon}). The observed coincidence of the jet break time with the change in absorbing column is definitively suggestive within this interpretation.

\section{Summary and Conclusions}

We have presented X-ray observations of GRB~190114C, from several tens of seconds to about 10 days. We find that during the observing time window the absorbing column to the source decreases by a factor of about two. A statistical analysis with a time-dependent absorption model shows that the variability cannot be well modeled as the result of photo-absorption of the medium along the line of the sight by the source photons. This result stems from the fact that the drop in the magnitude of $N_{\rm H}$ is observed at later times, after the most intense radiation has already passed through the medium. 
Any absorber that is close to the burst would be quickly photoionized and the observed column would decrease in the first few tens of seconds. Any absorber that is far from the burst would produce constant absorption. We could not find any configuration for which a progressively photoionized absorber can explain the late drop of the column density without an early, faster drop.

With the most straightforward interpretation not being supported by the data, we speculate on other possible physical mechanisms which may induce it. In particular, we argue that an absorber with a low-filling fraction can produce such an effect, and we derive the required properties of the absorber. The typical dimensions of the absorbing blobs are those of Eq.~2, $\sim 0.1$ pc. These are the typical sizes of ultra-compact HII regions or massive cores in molecular clouds.

One may also wonder why only a few GRBs showed changes in the absorbing column density. From an observational basis, time-dependent variable absorption is difficult to detect. Several high signal-to-noise X-ray spectra are needed, and these are not always available. A late-time decrease (or increase\footnote{ Ideally, if a blob enters the line of sight of the jet as the jet spreads out, one should observe an increase of the absorbing column density. This effect is possible but is even more difficult to observe.
This is because, as the blob enters the jet line of sight, only a fraction of the jet emitting area is affected, thus producing a small (partial) increase in $N_{\rm H}$.}) is even more difficult to detect as it requires good spectral data while the afterglow fades.

Another peculiarity of GRB~190114C is its connection with TeV emission. After the MAGIC detection \citep{GRB1901114}, two other firm GRB detections at TeV energies have been gathered: GRB~190829A (\citealt{deNaurois2019}) at $z=0.08$ (\citealt{Valeev2019}) detected by H.E.S.S., and GRB~201216C (\citealt{Blanch2020}) possibly at $z=1.10$ (\citealt{Vielfaure2020}) detected by MAGIC.
Also GRB~180720B at $z=0.65$ \citep{Vreeswijk2018} showed very high energy photons detected by H.E.S.S. \citep{Abdalla2019}. However, this emission came late ($\sim 11$ hr).

GRB~190829A was a bright burst but due to its proximity the overall energy is small $E_{\rm iso}\sim 2\times10^{50}$ erg and so the peak luminosity \citep{Tsvetkova2019}.
The burst was characterized by a high intrinsic column density ($N_{\rm H}\sim 2\times 10^{22}$ cm$^{-2}$), which showed signs of a slight decrease by $\sim 40\%$ within 10 ks.
GRB~201216C was a bright burst $E_{\rm iso}\sim 6\times10^{53}$ erg \citep{Frederiks2020} with a high intrinsic column density of $\sim 4\times 10^{22}$ cm$^{-2}$, too. Even though the {\it Swift} data span only a very narrow time interval (3-15 ks after the GRB onset), there is an indication for a column density decrease by a factor of $4.0\pm2.5$. For both bursts, there was no {\it Fermi} LAT detection. 
It is tempting to consider the high intrinsic column density as a common property of the TeV-emitting high- and low-luminosity GRBs. Such connection may be due to the fact that the TeV emission is likely coming from the external shock and a high density of the external material would affect both the absorbing column and the afterglow properties. The reason for a correlation between TeV emission and column variability is, however, more difficult to anticipate and a more detailed study is required to bring that out.

\begin{acknowledgements}
{The authors would like to thank the anonymous referee for their comments and suggestions which helped improving the quality of the manuscript. SC acknowledges support from the Italian Space Agency, contract ASI/INAF n. I/004/11/4. SC warmly thanks the {\it XMM-Newton} Project Scientist, Norbert Schartel, for approving Director's Discretionary Time observations. 
RP acknowledges support by NSF award
AST-2006839 and from NASA (Fermi) award 80NSSC20K1570. DL acknowledges support from NASA grant NNX17AK42G (ATP) and NSF grant AST-1907955.}
\end{acknowledgements}

\bibliographystyle{aa}
\bibliography{ms}

\end{document}